\documentclass[twoside,fleqn]{article}
\usepackage[dvips]{graphics}
\usepackage{espcrc2}
\usepackage{epsfig}
\usepackage{latexsym}

\def\lsim{\raise0.3ex\hbox{$<$\kern-0.75em\raise-1.1ex\hbox{$\sim$}}}
\newcommand{\bq}{\begin{equation}}
\newcommand{\eq}{\end{equation}}
\newcommand{\bqa}{\begin{eqnarray}}
\newcommand{\eqa}{\end{eqnarray}}
\newcommand{\bqas}{\begin{eqnarray*}}
\newcommand{\eqas}{\end{eqnarray*}}
\newcommand{\bdm}{\begin{displaymath}}
\newcommand{\edm}{\end{displaymath}}

\newfont{\mega}{cmr17 scaled 2000}
\newfont{\Mega}{cmr17 scaled 3000}
\newfont{\MEGA}{cmr17 scaled 4000}
\newfont{\giga}{cmr17 scaled 5000}
\newfont{\Giga}{cmr17 scaled 6000}
\newfont{\GIGA}{cmr17 scaled 8000}

\title{ The flavour and quark mass dependence of thermodynamic quantities
 in lattice QCD \thanks{ This work was partly supported by
    the Deutsche Forschungsgemeinschaft under grant Ka 1198/3-1 and the EU TMR
    network grant ERBFMRX-CT97-0122.}}

\author{ A. Peikert with F. Karsch, E. Laermann\\[3mm]Fakult\"at
    f\"ur Physik, Universit\"at Bielefeld, 33501 Bielefeld, Germany}

\begin{document}

\begin{abstract}
The dependence of thermodynamic properties of QCD on the number of
quark flavours is investigated. Lattice results for the equation
of state are presented for 2, 2+1 and 3 quark flavours.
The simulations have been performed with the improved p4-staggered
fermion and a Symanzik improved gluon action on lattices of size $16^3 \times 4$\
and $16^4$.
\end{abstract}

\maketitle

\section{INTRODUCTION}
In this paper thermodynamic observables like the critical
temperature $T_c/\sqrt{\sigma}$ and
the pressure $p/{T^4}$ have been calculated in full QCD with an improved
staggered fermion action. The use of improved actions in finite temperature
calculations is strongly suggested by results from the pure gauge sector
which have shown that improved actions reduce finite cut-off effects
in thermodynamic quantities significantly \cite{beinlich97,beinlich99}.
 In addition, the calculation
of the pressure with the standard staggered action did show strong cut-off
effects when comparing results obtained on lattices
with temporal extent $N_\tau=4$ and 6 \cite{bernard97}. \\
The relevant number of flavours in finite temperature QCD
is expected to be two light and one heavier quark. To quantify the effect
of varying the number of quark flavours simulations with $N_f=$2, 2+1 and 3
are performed.
\section{THE SIMULATION}
The fermion fields have been simulated using the tree-level p4 action
\cite{heller98} including fat-links \cite{blum97} with a fat-weight $\omega =
0.2$ in the one-link derivative. This action is constructed to improve the
rotational symmetry of the free quark propagator up to ${\cal
  O}(p^4)$. The high temperature ideal gas limit of the free energy is
approached much faster for the p4 than for the staggered action (Figure 1).
  \epsfig{file=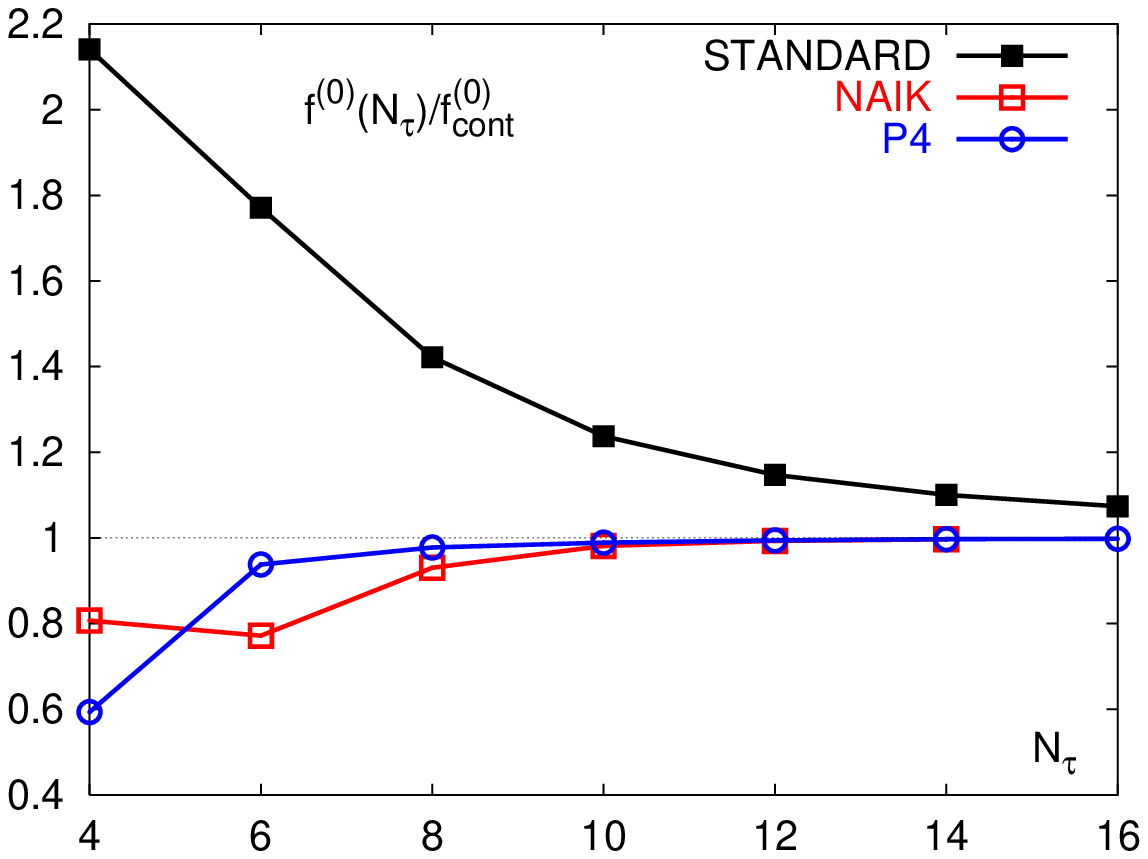,height=50mm,width=75mm}
{ Figure 1. The fermion free energy in the ideal gas limit. }\\[2mm]
Fat-links lead to an improved flavour
symmetry \cite{blum97} which results in a reduced pion splitting measured
by the quantity
\bdm
\delta= { m^2_{\pi_2}-m^2_\pi \over m^2_\rho}. \nonumber
\edm
In Figure 2, $\delta$ is shown for the staggered and p4 fat action measured
at different quark masses. Comparing results at the same lattice
{spacing $a$} one finds a reduction of the pion splitting of about a factor
of 2. \\
For the gauge fields the tree-level $1\times2$ action has been used which in
pure gauge simulations lead to an improved finite cut-off behaviour
for quantities like latent heat\cite{beinlich97} and pressure
\cite{beinlich99}. \\
The gauge and fermion fields have been updated with the standard Hybrid R algorithm with a
step size $\Delta \tau \lsim m_qa/2$ and a trajectory length of 0.8. The lattice
sizes are $16^3\times 4$ and $16^4$.
  \epsfig{file=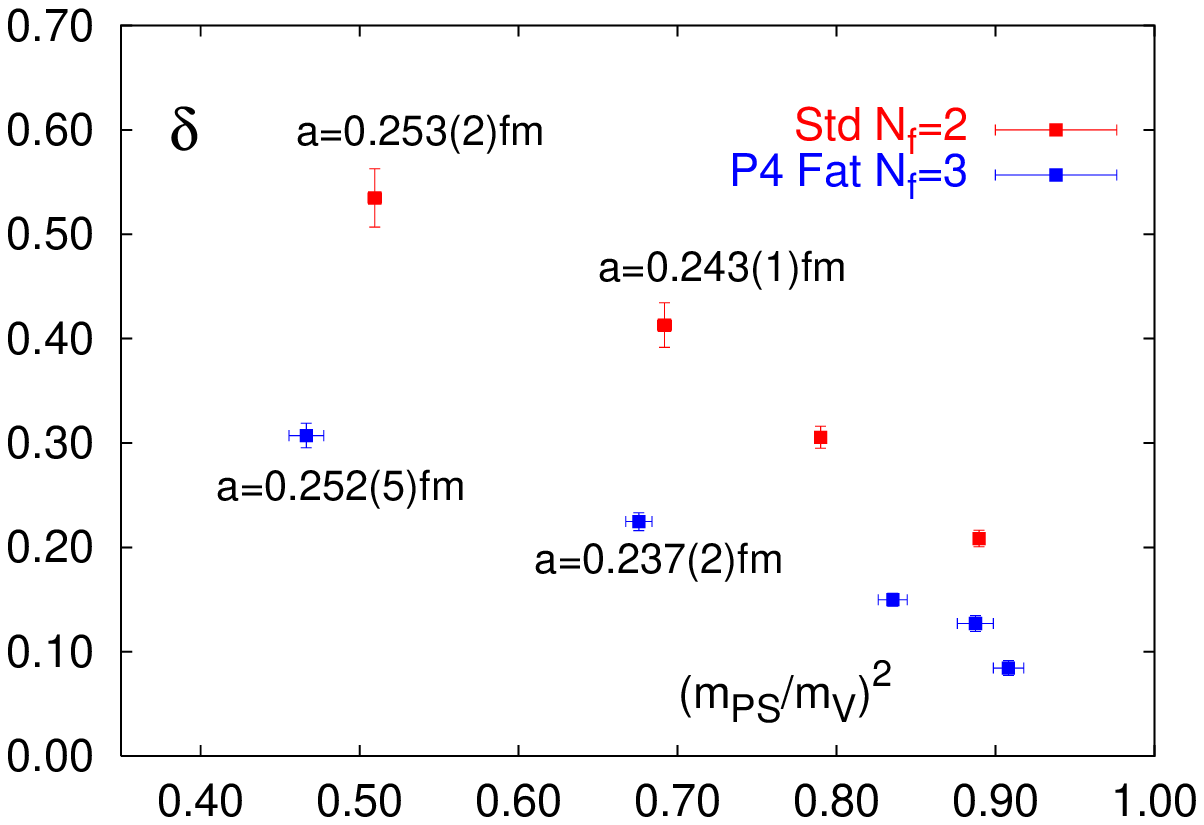,height=50mm,width=75mm}
{ Figure 2. The pion splitting $\delta$ for standard $(N_f=2)$ and
p4 fat improved $(N_f=3)$ actions. }\\[-4mm]

\section{THE CRITICAL TEMPERATURE AND THE TEMPERATURE SCALE}
The critical temperature has been calculated for various quark masses for 3
flavour QCD and in addition also for 2 and 2+1 flavours at a quark mass
of $m_q=0.10$. The pseudo-critical coupling $\beta_c$ was determined by the
peak position of the susceptibility of the Polyakov-loop and the chiral
condensate, respectively. At the pseudo-critical couplings zero-temperature
calculations on $16^4$ lattices have been performed measuring the
string-tension $\sqrt{\sigma}$ and the meson masses $m_{PS}$ and $m_{V}$. The
results for $T_c/\sqrt{\sigma}$ are plotted in Figure 3.
The qualitative behaviour of the 2 and 3 flavour values for the standard and p4
action is quite similar; in both cases $T_c$ drops quite fast already
\epsfig{file=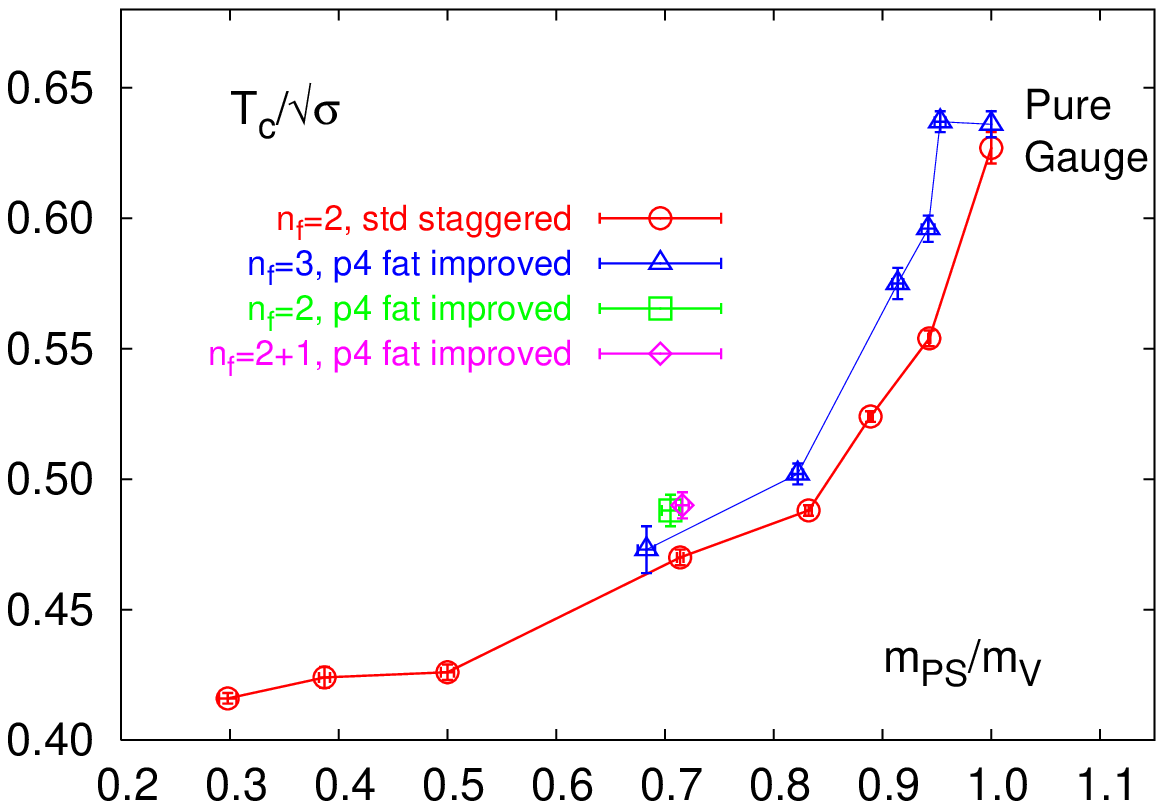,height=50mm,width=75mm}
{ Figure 3. The critical temperature $T_c/\sqrt{\sigma}$ for 2, 2+1 and 3
  flavours for the p4 action. For comparison also 2 flavour results
  obtained with the standard action \cite{luetgemeier98} are plotted.}\\[2mm]
at large
quark masses. For the improved p4 action $T_c/\sqrt{\sigma}$ for different
number of flavours does only show a small difference.\\
To set the temperature scale for the pressure calculations zero temperature
simulations have been performed at a variety of $\beta$ values for 2, 2+1 and 3
flavours. The resulting string tension data have been fitted to an ansatz
proposed by Allton \cite{allton96} (see Figure 4)
\bdm
 (a\sqrt{\sigma})(\beta)=R(\beta)(1+c_2\hat{a}^2(\beta)+c_4\hat{a}^4(\beta)+...)/c_0
\edm
with  $\hat{a}^2(\beta)={ R(\beta) / R(\beta_0)}$, the two-loop beta function
 $R(\beta)$ and the normalization $\beta_0$.\\
\epsfig{file=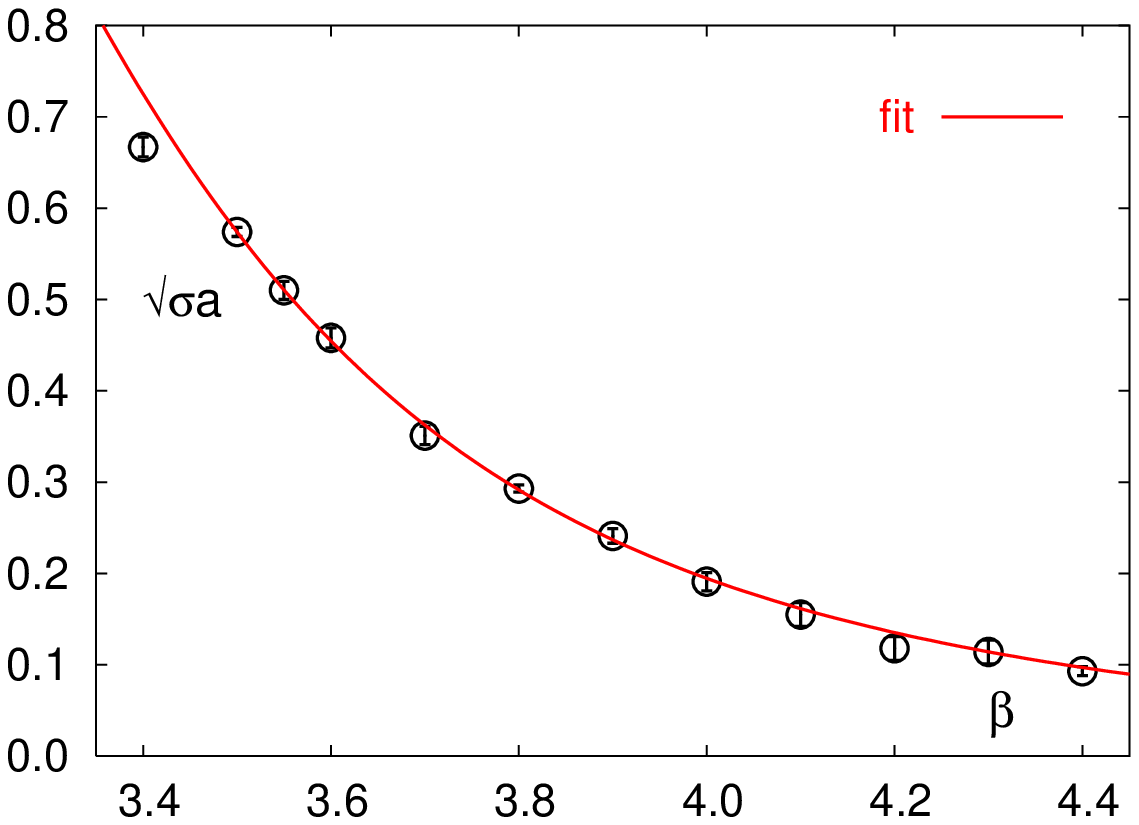,height=50mm,width=75mm}
{ Figure 4. The string tension data and the fit to the data for the $N_f=2+1$ case.}\\[-4mm]
\section{THE EQUATION OF STATE}
The pressure $p/T^4$ has been calculated for 2, 2+1 and 3
quark flavours. The masses of the light quarks are $m/T=0.4$, the mass of
the heavy quark is $m/T=1.0$.
From the difference of the gluonic part of the action
on finite temperature and zero temperature lattices
$\langle S^{1\times2}\rangle_0-\langle S^{1\times2}\rangle_T$ the pressure
can be calculated according to
\bdm
{p\over T^4}\bigg|^\beta_{\beta_{0}}=N_\tau^4 \int^\beta_{\beta_{0}} d\beta^\prime \left(\langle S^{1\times2}\rangle_0-\langle S^{1\times2}\rangle_T\right)
\edm
In Figure 5 the result for the differences of the gluonic action densities
are shown. One observes an increase for the maximum of that quantity
corresponding to the increase of the number of flavours. The integration
over these curves gives
 the pressure $p/T^4$ plotted in Figure 6. For larger
\epsfig{file=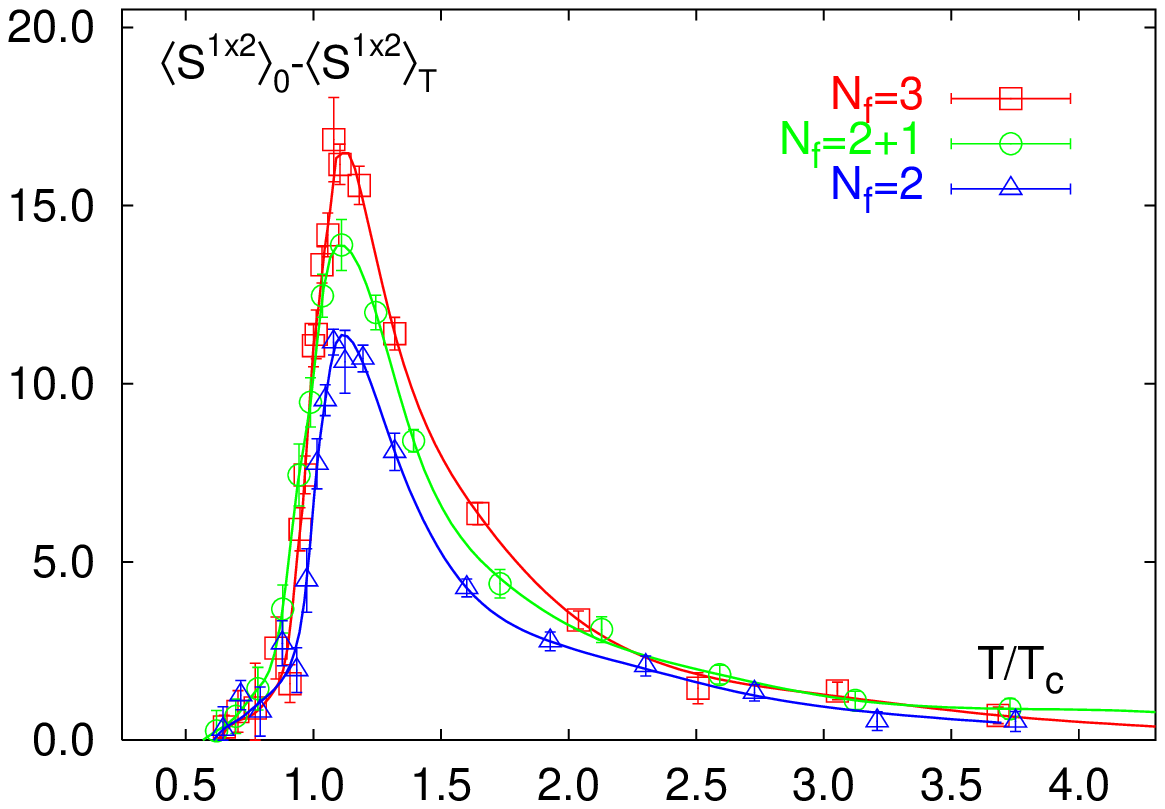,height=50mm,width=75mm}\\[-1mm]
{ Figure 5. The action differences $\langle S^{1\times2}\rangle_0-\langle
  S^{1\times2}\rangle_T$ for $N_f=2$, 2+1 and 3 for the p4 fat improved action.}\\[1mm]
temperatures the 2+1 flavour pressure is closer
to the 3 flavour than to the 2 flavour pressure
as expected from the high temperature Stefan Boltzmann limit (see arrows in Figure 6). If one normalizes
the pressure to the appropriate ideal gas value one finds
the same temperature
dependence for 2 and 3 flavours for the p4 action (Figure 7). Figure 7 also
shows that the systematic deviations from the Stefan
Boltzmann limit obtained with different fermion actions are qualitatively
in agreement with what has been calculated as the cut-off effect of the free
energy at $N_\tau=4$ for
the different actions (see Figure 1).\\
The cut-off effects of the 2 flavour pressure for the p4-fat action
and the standard staggered action are compared in Figure 8. As
the pressure calculated with the p4 action seems to scale with the
relevant number of degrees of freedom we may expect to obtain an
estimate for the
\epsfig{ file=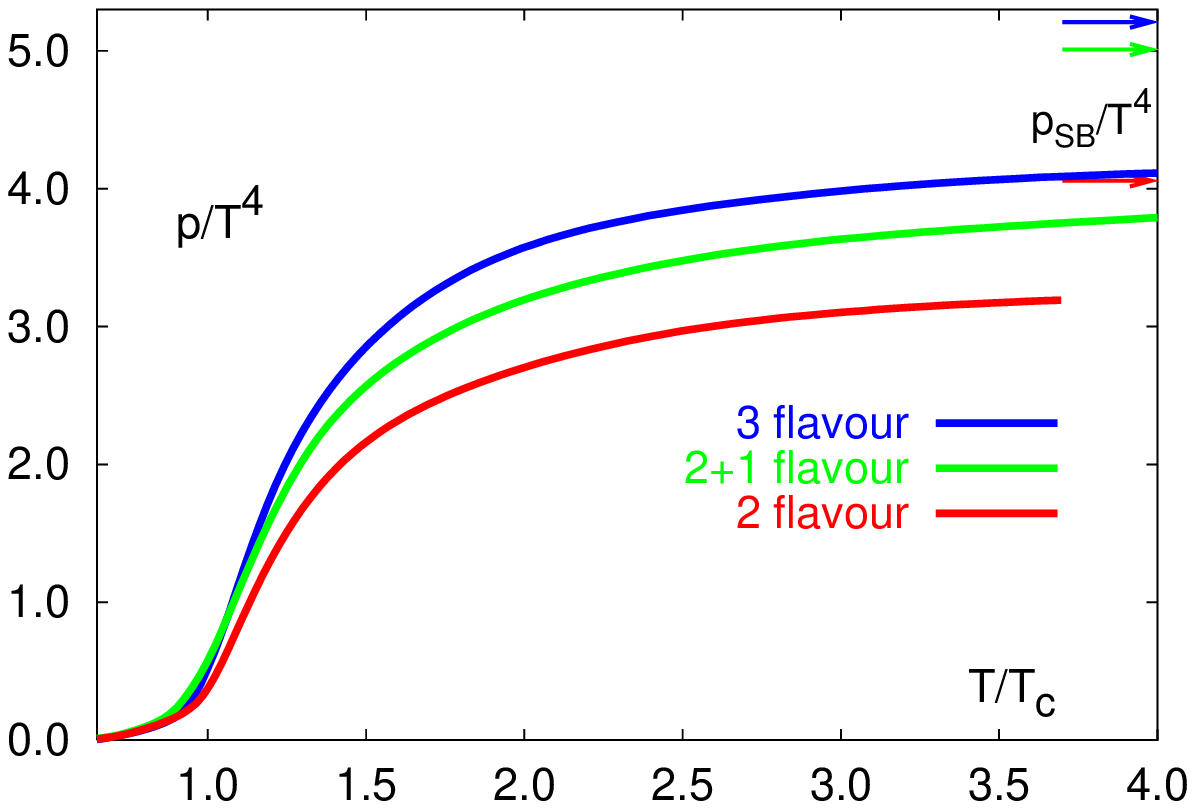,height=50mm,width=75mm}\\[-1mm]
{ Figure 6. The pressure for $N_f=2$, 2+1 and 3 for the p4 fat improved action.}
\epsfig{file=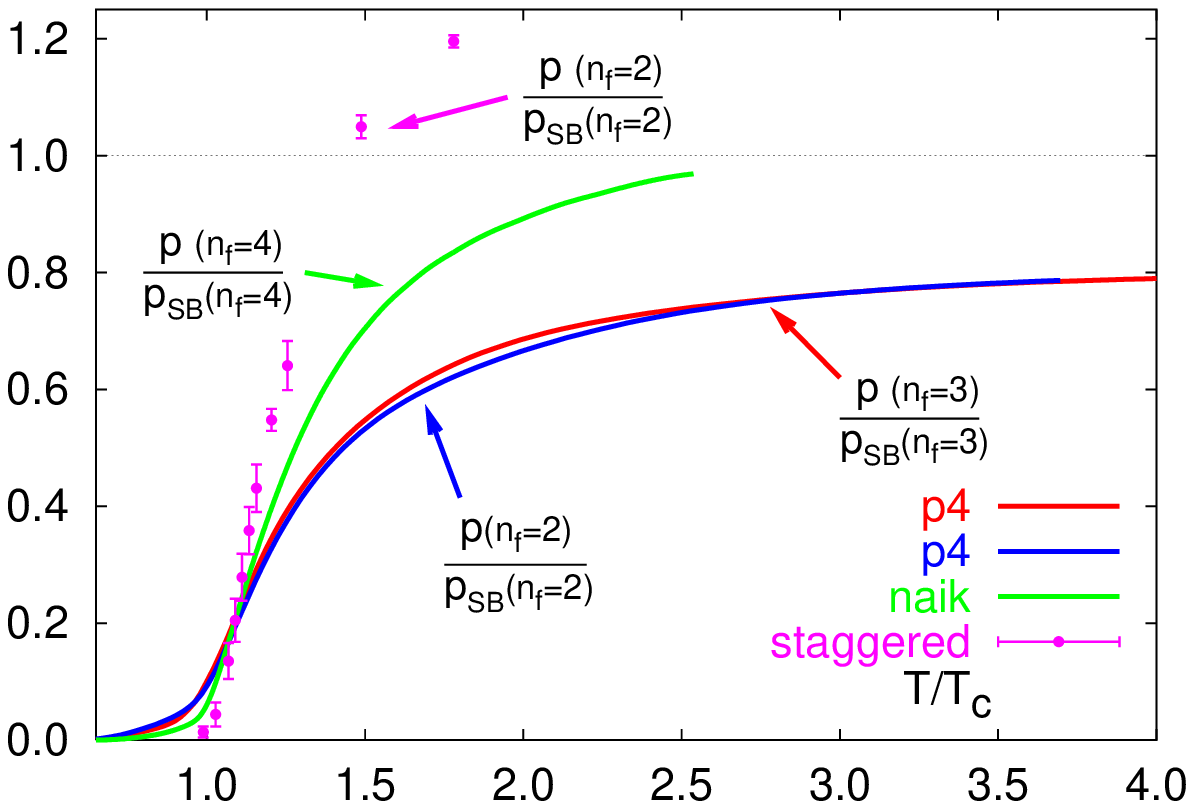,height=50mm,width=76mm}\\[-1mm]
{ Figure 7. The pressure normalized to the ideal gas value $p/p_{SB}$ for $N_\tau=4$. }\\[-1mm]
\epsfig{file=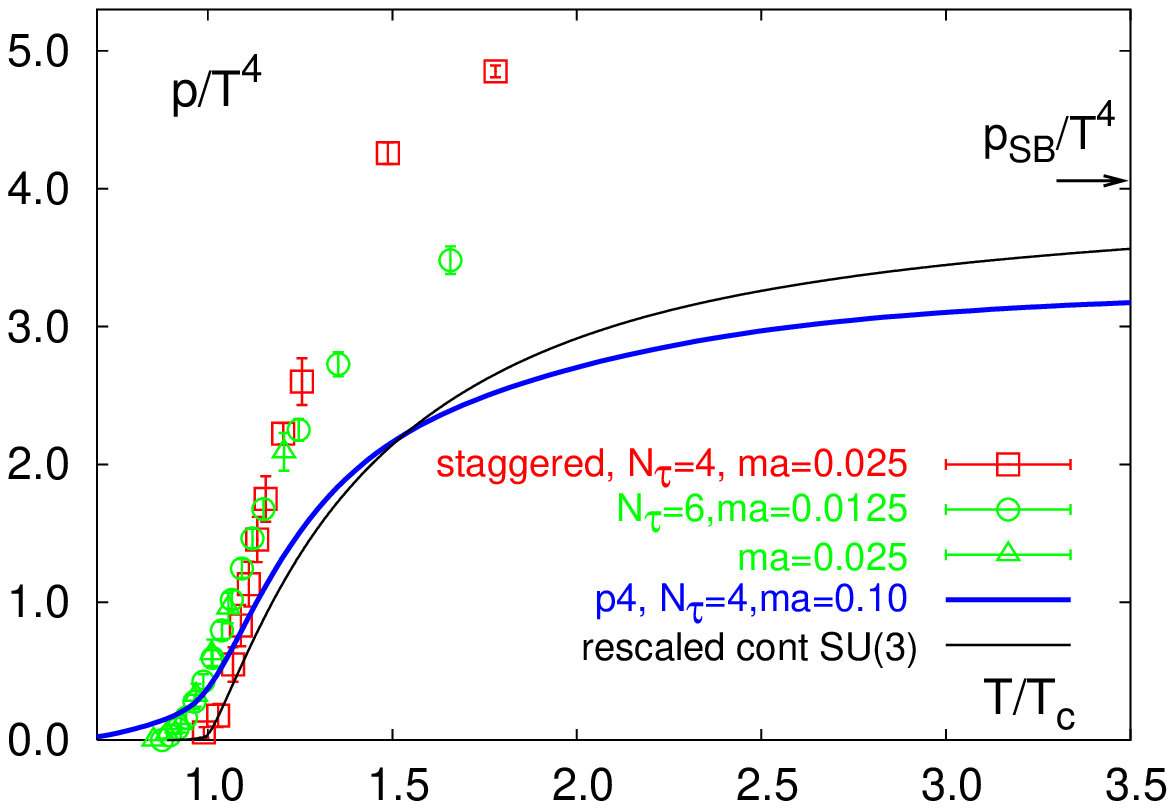,height=50mm,width=76mm}\\[-1mm]
{ Figure 8. The 2 flavour pressure for the standard  action (
  $N_\tau=4$, 6), the p4 fat action ($N_\tau=4$) and the rescaled
continuum pure gauge result. }\\[1mm]
 remaining cut-off dependence by comparing with the
appropriately rescaled continuum extrapolation
for the pure gauge pressure.
This is shown as the black curve in Figure 8. If this curve
approximates the continuum extrapolated pressure of full QCD at high
temperatures the p4 action indeed reduces strongly the cut-off effects.
A direct verification will require a simulation on a $N_\tau=6$ lattice.
\\[-6mm]

\end{document}